# Multifilament $YBa_2Cu_3O_{6+x}$ -coated conductors with minimized coupling losses.

G. A. Levin, P. N. Barnes, and J. W. Kell

Propulsion Directorate, Air Force Research Laboratory, 1950 Fifth St. Bldg. 450, Wright-Patterson Air Force Base OH 45433

N. Amemiya, Z. Jiang, K. Yoda, and F. Kimura

Faculty of Engineering, Yokohama National University,
79-5 Tokiwadai, Hodogaya, Yokohama 240-8501, Japan

**Abstract.**

*We report an experimental approach to making multifilament coated conductors with low losses in applied time-varying magnetic field. Previously, the multifilament conductors obtained for that purpose by laser ablation suffered from high coupling losses. Here we report how this problem can be solved. When the substrate metal in the grooves segregating the filaments is exposed to oxygen, it forms high resistivity oxides that electrically insulate the stripes from each other and from the substrate. As the result, the coupling loss has become negligible over the entire range of tested parameters (magnetic field amplitudes B and frequencies f) available to us.*

Currently, the main candidate for the broader commercialization of high temperature superconductor (HTS) wires is the second generation $YBa_2Cu_3O_{6+x}$ (YBCO)-coated conductor which is produced in the form of a wide, thin tape [1-3]. A major technical obstacle to their implementation in important power applications is the very large losses that occur in time-varying magnetic fields [4,5]. A way to reduce ac losses is to divide the superconducting layer into a large number of filaments (stripes) segregated by non-superconducting resistive barriers [6,7]. The defining factors for the commercial viability of the HTS coated conductor as a competitor to copper wire are the unit cost and the manufacturing throughput. It is therefore fundamentally important that the manufacturing process for the multifilament coated conductors was closely comparable in terms of economic factors with that of the standard HTS (uniform) coated conductors that are currently being developed.

Laser micromachining of coated conductors is a method that has been tested thus far more extensively than its potential alternatives (photolithography, etc.) [7-12]. In this approach, the superconducting layer of a uniform coated conductor is divided into parallel stripes by laser ablation. In the multifilament conductors, the amount of power dissipated in the superconducting layer (the hysteresis loss) is reduced in proportion to the width of an individual stripe. However, another channel of dissipation (coupling loss) appears as a result of the induced currents passing between the superconducting stripes through the normal barriers. During the ablation process the substrate metal melts and splashes across the walls of the newly created grooves segregating the superconducting stripes. The splashed metal provides the electrical connection between the stripes through the substrate resulting in very high coupling losses [8-12] It is well understood that a way to reduce coupling losses is to insulate the superconducting stripes from each other and from the substrate [9,13]. It was not clear, however, how to accomplish that in a way that would not negatively impact the rate of production of such a conductor and the cost. Here we present an approach to minimize the coupling losses that is effective and practical.

In this approach, we expose the striated YBCO-coated conductor to oxygen at a high temperature. The high resistivity oxides that form in the grooves effectively insulate the stripes from the bulk of the substrate. The measurements presented below indicate that after this post-ablation treatment the coupling loss is almost completely suppressed in the tested range of amplitude and frequency of the applied field.

The striated samples were made from a 4 mm wide uniform coated conductor provided by SuperPower Inc. [3]. The YBCO layer in this conductor was deposited on IBAD buffered Hastelloy substrate ~50 μm thick and covered with silver. Striation of the tapes is accomplished by laser micromachining utilizing a frequency tripled diode-pumped solid-state Nd:YVO$_4$ laser at a 355 nm wavelength. Details of striation by laser ablation are given in Refs. [9,11,14]. The 10 cm long, 4 mm wide sample was divided into 8 stripes, so that the width of an individual superconducting filament was close to 0.5 mm.

Figure 1 shows a profile of a groove separating the filaments.

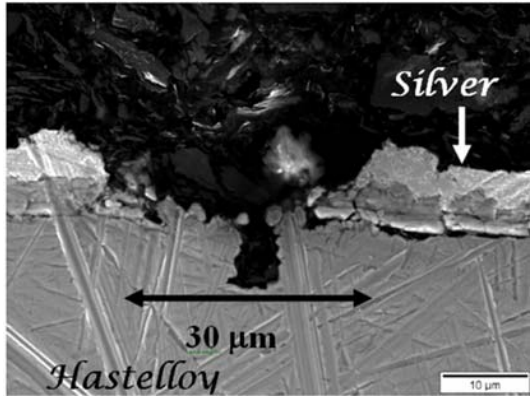

**Figure 1.** Profile of a groove; the depth is about 10 μm and it extends well into the Hastelloy substrate, cutting through silver, YBCO and buffer layers. The width of the groove is nominally close to 30 μm. The delamination from the substrate to the right may be the result of sample preparation for microphotography, rather than that of laser ablation.

Laser ablation creates rough, irregular shaped grooves in the substrate, completely removing the silver and YBCO from the exposed area. The width of the groove is about 30 microns although the damage to the YBCO layer might be more extensive due to its delamination from the substrate to the right of the groove. Delamination of this type occurs only in some areas and is more likely the result of preparing the sample for microphotography, rather than that of the laser ablation. The next step, that was not attempted previously [8-11], is to anneal the striated sample in flowing oxygen. The parameters of this procedure were deliberately kept similar to those required to oxygenate the YBCO in order to maximize its critical temperature. The temperature of the sample was raised gradually to $550^0$ C over 2-3 hours, kept at 500-$550^0$C for about 2-3 hours and reduced to room temperature over another 3 hours.

The losses in the striated as well as in the non-striated (control) samples were measured by placing them inside the bore of a dipole magnet that generated time-varying magnetic field as described in Ref. [8]. The total power loss per unit length in a rectangular multifilament coated conductor of width $W$ and length $L$ in the limit of full field penetration is determined by the sweep rate $Bf$, where $B$ is the amplitude of the applied field and $f$ is its frequency [4,11]:

$$Q = q_s Bf + q_n (Bf)^2 \ ; \ q_s \approx W_n I_c \ ;$$

$$q_n = \frac{\pi^2}{6} \frac{L^2}{R_{eff}} W \ . \quad (1)$$

The linear term describes the loss in the superconducting material and the quadratic term – the loss in the normal metal, predominantly the coupling loss. $W_n$ is the width of an individual filament and $I_c$ is the total critical current of the conductor. $R_{eff}$ is the phenomenological *coupling resistance* that characterizes the coupling loss.

Figure 2(a) shows the energy loss per cycle in a 10 cm long, 4 mm wide 8-filament sample after the post-ablation oxygenation. The range of frequencies and field amplitude is shown in the figure. The frequency dependence of $Q/f$ is weak which indicates that the main contribution to loss is the hysteresis loss. The value of the loss itself is not a very informative quantity because it can be reduced not only by filamentation of the superconducting layer, but also by the degradation of its critical current. A more objective measure, subject to minimization, is the amount of power loss per unit of transport critical current $Q/I_c$. The specific loss is defined as [10]

$$\Lambda \equiv \frac{Q}{I_c B f} = \lambda_1 + \lambda_2 (Bf); \quad \lambda_1 \approx W_n;$$

$$\lambda_2 = \frac{\pi^2}{6} \frac{L^2}{R_{eff} J_c}, \quad (2)$$

and it has the maximum value $\Lambda \to W_n$ when the coupling losses are negligible. Note, that both $\lambda_1$ and $\lambda_2$ do not depend on the total width of the conductor, except perhaps through $R_{eff}$. In Eq. (2) $J_c = I_c/W$ is the density of current per unit width. Therefore, the conductors of different widths can be objectively compared to each other on the basis of the specific loss.

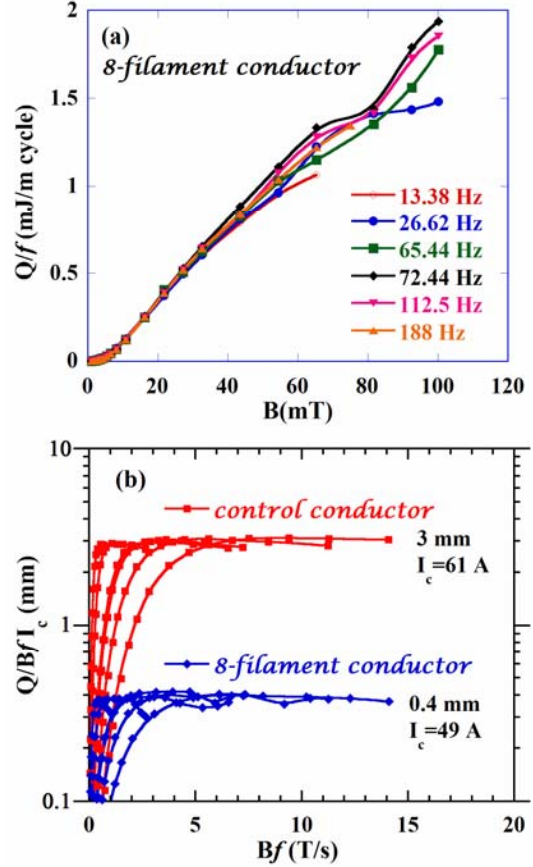

Figure 2: (Color online) (a) Energy loss per cycle in the 8-filament 100x4 mm$^2$ coated conductor vs amplitude of applied magnetic field. (b) The specific power loss vs sweep rate in the control, non-striated, and the 8-filament conductor. For both samples the data are shown for the same frequencies as in Fig 2(a). The loss is normalized by the respective critical currents indicated in the figure. The saturation levels of the specific loss, 3 mm and 0.4 mm respectively, are indicated.

In Fig. 2(b) we compare the specific loss in the 8-filament conductor (same data as in Fig. 2(a)) and a control (non-striated) 4 mm wide conductor. Both conductors were parts of a longer tape. The self-field critical current of the control sample is 61 A. Before striation the critical current in what will become the 8-filament sample was 55 A. After striation the critical current was reduced by 10% to 49 A. The data in Fig. 2(b) were normalized by the respective critical currents for either sample. When the field amplitude exceeds the penetration field (about *20 mT*) the specific loss saturates at the values *3 mm* for the control and *0.4 mm* for the multifilament sample (note that *1 mJ/T A m* $\equiv$ 1 mm). This corresponds to an almost 8-fold reduction, which is a maximum reduction of the specific loss that one can achieve in the 8-filament conductor. In the control sample one could expect the limit being $\Lambda \to W = 4mm$, but the hysteresis loss is determined by the effective width of the current-carrying layer which may be smaller than the physical width. Correspondingly, 0.4 mm is the effective width of an average filament.

An important point is that the coupling loss determined by the slope $\lambda_2$ ( Eq. (2)) is negligible in the 8-filament sample. In fact, it is so small that we cannot make a reliable estimate of $\lambda_2 = d\Lambda/d(Bf)$. This is in large contrast with the previous results obtained on the same type of multifilament conductors [10,11], but without post-ablation oxygenation. In Fig. 3 the specific loss for the 10 mm wide, 20-filament conductor [10] that was not oxygenated after ablation is shown along with that for the 8-filament conductor. Both have 0.5 mm separation between the grooves, which is why the specific hysteresis loss – the intercept $\lambda_1$ – is about the same ($\approx 0.45$ mm and 0.4 mm respectively). But the difference in the amount of coupling loss (the slope $d\Lambda/d(Bf)$) is obvious. In the samples without post-ablation oxygenation the value of the coupling resistance $R_{eff}$ is close to the sheet resistance of the substrate $R_{eff} \approx \rho/d_n$, where $\rho$ and $d_n$ is the resistivity and the thickness of the substrate, respectively [10,11]. Apparently, after oxygenation the coupling resistance increases by at least two orders of magnitude.

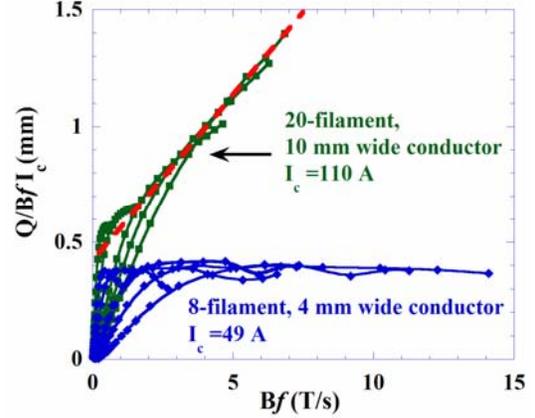

Figure 3 (Color online) Specific loss of the 8-filament sample, Fig. 2(b), and that of the 20-filament, 10 mm wide sample that was not subjected to post-ablation oxygenation. The dashed straight line is a linear fit that corresponds to Eq. (2). The slope is the measure of the coupling losses.

These results raise two main questions: (a) the nature of the chemical reactions initiated by the post-ablation annealing and (b) the reason why this treatment leads to such dramatic increase in coupling resistance. Hastelloy is mostly a Ni-Cr-Mo-W alloy and the exposure of the grooves created by ablation, Fig.1, to oxygen at high temperature leads to the formation of high resistivity oxides such as NiO, $Cr_2O_3$, etc. or compound oxides such as spinels $NiO \cdot Cr_2O_3$ that form at higher temperatures. The presence of molybdenum increases the rate of oxidation which is attributed to the volatility of the oxide $MoO_3$ preventing the formation of a stable protective scale [15]. A likely more important factor accelerating the kinetics of oxidation is the highly disrupted and rugged structure of metal left after laser ablation. Owing to microscopic cracks, oxygen can penetrate deeper into the substrate through the large exposed surface area and nucleate oxides at favorable sites, such as dislocations and impurity atoms.

Non-metallic, high resistivity oxides encapsulate the superconducting stripes (as well as the Ag cap layer on top of them) at the edges, and at low temperature practically insulate them (together with the insulating buffer) from the metallic substrate. Although we cannot determine how deep into the substrate the oxide barriers intrude, it is unlikely that they penetrate all the way to the bottom. Therefore, the tremendously strong reduction of the coupling loss indicate that the transverse electric field is concentrated close to the surface of the groove [16].

In summary, we have demonstrated that the multifilament coated conductors with very low coupling losses can be obtained by laser ablation with subsequent oxygen annealing. The annealing parameters, temperature and duration, can be the same as those required for oxygenation of YBCO.

**Acknowledgements:** We thank T. Campbell for technical assistance. At YNU this work was partially supported by the US Air Force Office of Scientific Research under contract AOARD-03-4031. One of the authors, G.A.L, was supported by the National Research Council Senior Research Associateship Award at the Air Force Research Laboratory.